\documentclass[prl,floatfix,shwopacs,amsmath,twocolumn]{revtex4}
\usepackage{amssymb}
\usepackage{graphics}
\begin{document}
\newcommand{\one}{\ensuremath{\hbox{$\mit I$\kern-.3em$\mit I$}}}
\renewcommand{\one}{\ensuremath{\hbox{$\mathrm I$\kern-.6em$\mit 1$}}}
\renewcommand{\one}{\ensuremath{\hbox{$\mathrm I$\kern-.6em$\mathrm 1$}}}
\newcommand{\be}{\begin{equation}}
\newcommand{\ee}{\end{equation}}
\newcommand{\bea}{\begin{eqnarray}}
\newcommand{\eea}{\end{eqnarray}}
\newcommand{\bma}{\begin{subequations}}
\newcommand{\ema}{\end{subequations}}
\def\qed{\leavevmode\unskip\penalty9999 \hbox{}\nobreak\hfill
     \quad\hbox{\leavevmode  \hbox to.77778em{%
               \hfil\vrule   \vbox to.675em%
               {\hrule width.6em\vfil\hrule}\vrule\hfil}}
     \par\vskip3pt}

\title{Quantum nonlocality in the presence of superselection rules and
data hiding protocols}
\author{F. \surname{Verstraete}}
\affiliation{Max-Planck-Institut f\"ur Quantenoptik,
  Hans-Kopfermann-Str. 1, Garching, D-85748, Germany.}
\author{J.I. \surname{Cirac}}
\affiliation{Max-Planck-Institut f\"ur Quantenoptik,
  Hans-Kopfermann-Str. 1, Garching, D-85748, Germany.}
\date{February 5, 2003}
\begin{abstract}
We consider a quantum system subject to superselection rules, for
which certain restrictions apply to the quantum operations that
can be implemented. It is shown how the notion of
quantum--nonlocality has to be redefined in the presence of
superselection rules: there exist separable states that cannot be
prepared locally and exhibit some form of nonlocality. Moreover,
the notion of local distinguishability in the presence of
classical communication has to be altered. This can be used to
perform quantum information tasks that are otherwise impossible.
In particular, this leads to the introduction of perfect quantum
data hiding protocols, for which quantum communication (eventually
in the form of a separable but nonlocal state) is needed to unlock
the secret.
\end{abstract}
\maketitle

The laws of Quantum Mechanics allow us to carry certain tasks that
would be impossible in a classical world. The strength of quantum
information theory resides in a subtle interplay between the
grandeur of the Hilbert space and the limitations on the allowed
operations. One of the most prominent examples is quantum
cryptography \cite{BB84}, where the existence of quantum
superpositions or entangled states on one hand and the limitations
due to the no-cloning theorem \cite{WZ82} on the other hand ensure
the possibility of a secure transmission of information between
two or more partners.

The power of quantum mechanics becomes especially apparent when
natural limitations apply to the operations that one can apply. As
an example, the theory of entanglement and quantum nonlocality
arises from the restriction to local operations and classical
communication (LOCC). It is somehow expected that additional
constraints would lead to new interesting physics and
applications, especially in the context of cryptography.

In many physical systems of interest, such an additional
restriction applies in form of superselection rules: an extra
\emph{axiom} of quantum mechanics dictates the existence of
superselection rules forbidding e.g. coherent superpositions
between states of elementary particles with different electric
charge \cite{WWW52,GP}. In practice, energy constraints lead to
effective superselection rules. We will consider a particular
instance of a superselection rule, which is deeply motivated by
current quantum optical experiments, namely particle number. This
corresponds, for example, to the situation in which one has
bosonic atoms as it is the case in experiments with cold atomic
gases \cite{BEC}. In this case, all physical observables (and
hence states) commute with the particle number operator. Our
results however also apply other types of superselection rules.

In this paper we will show that in the presence of superselection
rules it is possible to carry out tasks which otherwise cannot be
performed in the context of quantum information. The underlying
cause appears to be that superselection rules completely alter the
notion of local distinguishability of quantum states: it is not
true anymore that any two pure orthogonal states can be
distinguished using LOCC, as is always possible without
superselection rules \cite{WSH00}. We will exploit this fact to
construct a data hiding protocol \cite{datahid,datahid2}, where
the goal is to distribute some information (classical or quantum)
among several partners in such a way that they can only "read" it
if they are provided with the means to perform joint measurements.
It has been proven that a perfect scheme is impossible
\cite{datahid}. In the presence of superselection rules however,
we will show that perfect data hiding is possible, and that one
can do it with pure states.

Furthermore, we will show that our view of entanglement and
quantum nonlocality has to be altered \cite{Zan01,Barn,Wiseman}.
Entangled states are usually defined as those which cannot be
prepared by LOCC out of a product state. This definition
highlights the fact that in order to entangle two systems they
must interact with each other (eventually via some third system).
It distinguishes these states from those that are classically
correlated (equivalently, separable), i.e., which can be written
in the form \cite{Wer89}
 \be
 \label{sep}
 \rho =\sum_k p_k |a_k\rangle\langle a_k| \otimes |b_k\rangle\langle
 b_k|,
 \ee
where $1\ge p_k\ge 0$ and $|a_k\rangle$ and $|b_k\rangle$ are
normalized states. When we are dealing with superselection rules
however, it happens that separable states cannot necessarily be
prepared locally anymore, giving rise to the existence of states
that are separable but nonlocal. We will show how this gives rise
to a new type of nonlocal resource, as these states can be used to
act as mediators to implement operations that are impossible by
LOCC.

Finally, we will also show that all quantum information protocols
can also be implemented in the presence of such superselection, so
that these rules in practice do not impose any restriction to what
one can do in practice, but rather offer novel ways to implement
protocols that otherwise may not be possible. Note however that,
unfortunately, the impossibility proof of quantum bit commitment
\cite{bc} remains valid \cite{May02}.


We start out considering a set of particles and the corresponding
Hilbert space $H$. We can always decompose
 \be
 H=\oplus_{N=0}^\infty H_N
 \ee
where $H_N$ is a subspace with a total number $N$ of particles. We
assume that the particle number is a superselection observable, in
the sense that the corresponding operator commutes with all
observables \cite{GP}. This immediately imposes that superposition
of pure states with different particle number cannot be prepared.
This is the case, for example, in all the experiments with atoms
or electrons. Any density operator must therefore admit a
decomposition of the form
 \be
 \label{rhoss}
 \rho= \sum_{N=0}^\infty p_N \rho_N,
 \ee
where $\rho_N$ is supported in $H_N$.

The situation becomes more intriguing when we consider two
systems, A and B, spatially separated. Then, $H_N=\oplus_{n=1}^N
(H^A_n\otimes H^B_{N-n})$, where $H^A_n$ ($H^B_n$) denotes a
Hilbert space corresponding to system A (B), with $n$ particles.
The superselection rule combined with locality imposes that $A$
and $B$ cannot prepare superposition states of different local
number of particles by LOCC. As we now show, this has very deep
consequences in the concept of entanglement. In fact, there might
be states which are separable but still they cannot be prepared
locally, and therefore they are nonlocal. Let us consider two
simple examples of such states. We take the simplest case in which
$H^{A,B}$ are one--dimensional, i.e. they are spanned by the
vectors $|n\rangle_{A,B}$ with $n$ particles.

Example 1:
 \be
 \label{Ex1}
 \rho_1=\frac{1}{4}(|0\rangle_A\langle 0| \otimes |0\rangle_B\langle 0|
 + |1\rangle_A\langle 1| \otimes |1\rangle_B\langle 1|)
 +\frac{1}{2}|\Psi_+\rangle_{AB}\langle \Psi_+|,
 \ee
where $|\Psi_+\rangle_{AB}=(|0\rangle_A|1\rangle_B+
|1\rangle_A|0\rangle_B)/\sqrt{2}$. This state is separable and has
a very simple separable decomposition with $p_k=1/4$ ($k=1,2,3,4$)
and
 \bma
 \bea
 |a_{1,2}\rangle &=& |b_{1,2}\rangle:=\frac{1}{\sqrt{2}}(|0\rangle \pm
 |1\rangle),\\
 |a_{3,4}\rangle &=& |b_{3,4}\rangle:=\frac{1}{\sqrt{2}}(|0\rangle \pm i
 |1\rangle).
 \eea
 \ema
Note that all these states are not compatible with the local
version of the superselection rule, since they involve a
superposition of different number of particles, and this applies
to all separable decompositions.

Example 2:
 \be
 \label{Ex2}
 \rho_2 =
 \int_0^{2\pi} \frac{d\phi}{2\pi}
 |\alpha e^{i\phi}\rangle_A\langle\alpha e^{i\phi}|\otimes
 |\alpha e^{i\phi}\rangle_B\langle\alpha e^{i\phi}|,
 \ee
where $\alpha>0$ and
 \be
 \label{coh}
 |\alpha e^{i\phi}\rangle:= e^{-\alpha^2/2} \sum_{n=0}^\infty
 \frac{\alpha^n}{\sqrt{n!}} e^{i\phi n} |n\rangle,
 \ee
denotes a coherent state. This state can be written in the form
(\ref{rhoss}) since it commutes with the total number operator,
and therefore is compatible with the superselection rule. On the
other hand, this state is obviously separable, though the states
$|\alpha e^{i\phi}\rangle$ (\ref{coh}) are incompatible with the
superselection rule.

Let us now show that the states (\ref{Ex1},\ref{Ex2}) cannot be
prepared locally if superselection rules apply. We will first
derive a general result which is not only valid for the simple
case in which $H^{A,B}_n$ are one--dimensional. Thus, these
subspaces could now have an arbitrary number of dimensions, which
include, for example, the use of ancilliary systems, and other
modes or degrees of freedom of the particles. We just need to
define $P^A_n$, the projector onto $H^A_n$ and analogously for
$P^B_n$.

{\bf Proposition 1:} If $\rho$ can be prepared locally, then
 \be
 \label{cond_local}
 \rho={\cal N}(\rho):=\sum_{n_A,n_B=0}^\infty (P^A_{n_A}\otimes P^B_{n_B})\rho
 (P^A_{n_A}\otimes P^B_{n_B}).
 \ee
{\it Proof:} $\rho$ can be written as a convex combination of
$|a_{n_A}\rangle_A\otimes|b_{n_B}\rangle$ with $|a_{n_A}\rangle\in
H^A_{n_A}$ and $|b_{n_B}\rangle\in H^B_{n_B}$ which themselves
fulfill (\ref{cond_local}). One can easily show that ancilliary
systems do not affect this property.\qed

Coming back to Examples 1 and 2, we have
 \bma
 \bea
 {\cal N}(\rho_1) &=& \frac{1}{4} \sum_{n,m=0}^1 |n\rangle_A\langle
 n|\otimes |m\rangle_A\langle m| \ne \rho_1,\\
 {\cal N}(\rho_2) &=&
 \int_0^{2\pi} \frac{d\phi_1}{2\pi} |\alpha e^{i\phi_1}\rangle_A\langle \alpha e^{i\phi_1}|
 \nonumber\\ &\otimes&
 \int_0^{2\pi} \frac{d\phi_2}{2\pi}
 | \alpha e^{i\phi_2}\rangle_B\langle \alpha e^{i\phi_2}|\ne \rho_2.
 \eea
 \ema
Thus, as announced above, both $\rho_{1,2}$ are separable states
which cannot be locally prepared and are therefore expected to
exhibit some kind of nonlocal properties.

The dual problem to the local preparation of quantum states is the
problem of locally distinguishing quantum states (eventually with
the help of classical communication). In the presence of
superselection rules, the following applies:

{\bf Proposition 2:} The states $\rho$ and ${\cal N}(\rho)$ cannot
be distinguished using local operations and classical
communication (LOCC).

{\it Proof:} First, note that we do not need to consider POVMs
since we can always include the state of the ancillas in $\rho$.
The operator corresponding to any observable $X_A$ ($X_B$) that
Alice (Bob) measures has to commute with her (his) particle number
operator, and thus we can write ${\rm tr}[(X_A\otimes X_B)
\rho]={\rm tr}[{\cal N}(X_A\otimes X_B) \rho]={\rm tr}[(X_A\otimes
X_B) {\cal N}(\rho)]$.\qed

Now, let us turn to introduce a quantum information task which
makes use of these ideas. Since the local distinguishability is
drastically affected by the presence of superselection rules, it
is natural to investigate quantum data hiding protocols
\cite{datahid,datahid2}. We will show how a third party can give a
secret bit to Alice and Bob which cannot be disclosed if they are
only allowed to use local operations and classical communication
(LOCC). If the bit is $0$ or $1$, the state
 \be
 |\pm\rangle:=\frac{1}{\sqrt{2}} (|0\rangle_1|1\rangle_2 \pm|1\rangle_1|0\rangle_2),
 \ee
is prepared, respectively. Then system 1 is given to Alice whereas
system 2 to Bob. Note that the states $|\pm\rangle$ contain a
superposition of two one--particle states and can therefore be
prepared by the third party. In order to show that Alice and Bob
can get no information about the bit, it is sufficient to note
that ${\cal N}(|+\rangle\langle +|)={\cal N}(|-\rangle\langle
-|)$: proposition 2 ensures that Alice and Bob cannot learn the
value of the bit, even if they use classical communication. This
value can be obviously read if they are allowed to perform joint
operations. Notice that the scheme is perfect and uses pure
states, in contrast to what happens in the scenario without
superselection rules  where perfect data hiding is not possible
\cite{datahid}. Note also that following \cite{datahid}, the
present scheme can be used to hide quantum bits.

Next, we analyze the resources needed by Alice and Bob to learn
the value of the bit if they can perform LOCC and they share
entanglement. First, let us assume that they are given the
entangled state
 \be
 |\Psi\rangle_{AB}= \frac{1}{\sqrt{N+1}} \sum_{n=0}^N |n\rangle_A
 |N-n\rangle_B.
 \ee
Thus, the total state will be
 \begin{eqnarray*}
 |\Psi_\pm\rangle &\propto&
 |0,0\rangle_A|1,N\rangle_B \pm
 |1,N\rangle_A|0,0\rangle_B \\
 &&\hspace{-1.2cm} + \sum_{n=1}^{N}
 (|0,n\rangle_A|1,N-n\rangle_B
 \pm |1,n-1\rangle_A|0,N-n+1\rangle_B).
 \end{eqnarray*}
We consider the following local measurement. Alice and Bob measure
both in the orthonormal basis composed of
 \be
 |\pm,n\rangle:=\frac{1}{\sqrt{2}}(|0,n\rangle\pm |1,n-1\rangle)
 \ee
($n>0$) and $|+,0\rangle:=|0,0\rangle$,
$|-,0\rangle:=|1,N\rangle$. Then, each of them assigns the value 0
(1) to the measurement if the outcome of the measurement
corresponded to one of the states $|+,n\rangle$ ($|-,n\rangle$)
for some $n$. In case both assignments are the same (different),
then they infer that the hidden bit was $0$ ($1$). One can easily
see that they will guess the value of the bit with a probability
$N/(N+1)$. We see that the probability is smaller than one but
approaches this value for $N\to\infty$.

It seems to be a distinctive feature of distributed quantum
systems subject to superselection rules that no perfect
discrimination is possible in the presence of a bounded amount of
nonlocal resources: unlike the usual case where teleportation can
be used to create a quantum channel through a classical channel
assisted by entanglement, this is in general not possible anymore
in the presence of superselection rules. A classical channel
assisted by a finite amount of entanglement is not equivalent
anymore to a quantum channel \cite{IP}. In particular, this
implies that our quantum data hiding scheme can be made arbitrary
secure in the presence of a bounded amount of nonlocal resources
shared between Alice and Bob \cite{note2}.

Now we show that the separable states that cannot be prepared
locally can also help to reveal the value of the hidden bit. This
indeed proves that they are "more useful" to perform certain tasks
than the ones that can be prepared locally, and therefore give
rise to a new kind of nonlocal resource \cite{IP}. We consider
that, apart from the shared state, they are given the separable
state $\rho_2$ [Eq.(\ref{Ex2})]. Then, they perform the same local
measurement as before and choose the value of the bit in the same
way. Let us denote by $P^z_{x,y,n,m}$ the probability that they
obtain the outcomes corresponding to $|x,n\rangle$ and
$|y,m\rangle$, respectively, if the bit was $z$ ($x,y=\pm$ and
$z=0,1$). We have
 \bea
 f_{n,m}(\alpha)&=&P^0_{+,+,n,m}=P^0_{-,-,n,m}=P^1_{+,-,n,m}=P^1_{-,+,n,m}\nonumber\\
 &=&\frac{e^{-2\alpha^2}}{4}\frac{|\alpha|^{2(n+m-1)}}{n!m!}
 |\sqrt{n}+\sqrt{m}|^2.
 \eea
One can readily prove that $\sum_{n,m=1}^{\infty}f_{n,m}(\alpha)$
tends to 1 in the limit $\alpha\to \infty$. Thus,  the probability
of detecting the value of the bit can be made arbitrarily close to
one without using a non--separable state, although we have proven
that this was not possible with LOCC operations.

Let us next formulate a few extensions to the present quantum data
hiding scheme. First of all, it can easily be shown that Alice and
Bob cannot extract any information even if they are given multiple
copies of the same hiding state: this follows from the fact that
${\cal N}(\rho_1^{\otimes M})={\cal N}(\rho_2^{\otimes M})$ for
any number of copies $M$. This fact could be very useful when
quantum data hiding would be implemented in a realistic
environment with decoherence. Secondly, our scheme can readily be
generalized to the multipartite setting \cite{datahid2}. Consider
for example the following $N$-party state:
\[
|\pm\rangle:\propto|0\rangle_1|1\rangle_2\cdots|N-1\rangle_N\pm
|1\rangle_1|2\rangle_2\cdots|N\rangle_{N-1}|0\rangle_N \]

One immediately sees that this hiding scheme is perfectly secure,
even when $N-1$ parties would decide to join forces. Note that
more sophisticated versions can readily be constructed.

So far we have shown that the particle number superselection rules
restricts the action that can be performed locally, which may be
used to perform certain tasks that otherwise would be impossible.
This seems to suggest that such rules may restrict some quantum
information protocols. Now we show that this is not the case, i.e.
that it is always possible to perform such protocols. The idea is
quite simple and consists of noticing that we can always consider
states that have a fixed number of local particles. Thus, if we
want to have a protocol using a $d+1$--level systems in one
particular location, we can just take states of the form
$|n\rangle:=|n,d-n\rangle$ ($n=0,1,\ldots,d$) in that location.
Obviously, superselection rules do not give any restriction in the
manipulation of these states. More specifically, a genuine qubit
can be encoded in the subspace spanned by the states
$|01\rangle,|10\rangle$, and in this case one can readily verify
that all known normal quantum information tasks such as
teleportation and quantum error correction can be implemented.

Finally, we discuss a possible set--up where our ideas can be
physically implemented and proof of principle experiments may be
carried out. We propose to use atoms as particles, since the
number of atoms can be considered as a superselection rule. Let us
consider, for simplicity, a set of bosonic atoms, each of them
with two internal (ground) levels $|a\rangle$ and $|b\rangle$. We
will assume that they are at very low temperature, as it is
usually achieved in Bose--Einstein condensation experiments
\cite{BEC}. We will also ignore the effect of interactions,
something which can be achieved by appropriately tuning the
scattering length. We will denote by $a_0$ ($b_0$) the
annihilation operators of atoms in internal state $|a\rangle$
($|b\rangle$) and motional state $|\psi_0\rangle$, the one
corresponding to the ground state of the Bose--Einstein
condensate. Denoting by $N$ the initial number of atoms, the
initial state can be written in second quantization as
  \be
  |\Psi(0)\rangle= \frac{(a_0^\dagger)^N}{\sqrt{N!}} |{\rm vac}\rangle.
  \ee
If a pair of laser in Raman configuration is applied for the
appropriate time, the state will be the same as before but with
$a_0 \to (a_0 \pm b_0)/\sqrt{2}$, where the sign $\pm$ can be
easily adjusted with the laser phase--difference and it is chosen
according to the value of the bit that needs to be hidden. We will
assume that the trap holding the atoms in the internal state $a$
can be manipulated independently of the one for atoms in $b$
\cite{Jetal}. Thus, they can now adiabatically be spatially
separated, so that the atoms remaining in state $|a\rangle$ are in
a different location than those in $|b\rangle$. The first are
given to Alice, where the second are given to Bob. One can readily
see that the sign $\pm$ cannot be read by using local measurement
and classical communication, something which is due to the
conservation of the atom number. In order to read the result, one
can bring the atoms back to the initial situation and measure the
internal atomic state by applying a laser pulse that transforms
$a_0 \to (a_0 + b_0)/\sqrt{2}$ and $b_0 \to (a_0 - b_0)/\sqrt{2}$.
If the atoms are found in state $|a\rangle$ the bit was zero and
otherwise it was one. Let us emphasize that a Bose--Einstein
condensate is not required to perform this experiments, and that
even the typical atom interferometry experiments already show the
effect that we are discussing \cite{Myln}. On the other hand, the
number of atoms $N$ does not need to be known. Moreover, the
readout of the state of the bit can be accomplished by using
another Bose--Einstein condensate which is split again into two
parts but now with a well defined value of the sign. An
interference experiment in each of the sides will then reveal the
value of the bit \cite{CGZ}. Finally, let us remark that, in
practice, one can also perform the experiments with a simple
set--up which uses laser light instead of atoms.

\acknowledgments We thank Sandu Popescu for triggering our
interest in superselection rules and quantum information. This
work was supported in part by the E.C. (RESQ IST-2001-37559) and
the Kompetenznetzwerk "Quanteninformationsverarbeitung" der
Bayerischen Staatsregierung.


\begin{thebibliography}{99}
\bibitem{BB84} {C.H. Bennett and G. Brassard, Proc. of the IEEE Int. Conf. on
              Comp., Systems and Signal Proc., 175
              (1984).}
\bibitem{WZ82} {W.K. Wootters and W.H. Zurek, Nature \textbf{299},
802 (1982).}

\bibitem{WWW52} {G.C. Wick, A.S. Wightman and E.P. Wigner, Phys.
Rev. \textbf{88}, 101 (1952).}

\bibitem{GP} {A. Galindo and P. Pascual, {\em Quantum Mechanics I}, Springer-Verlag, Berlin (1990).}

\bibitem{BEC} {E.A. Cornell and C.E. Wieman, Rev. Mod. Phys. \textbf{74}, 875 (2002);
W. Ketterle, Rev. Mod. Phys. \textbf{74}, 1131 (2002).}

\bibitem{WSH00} {J. Walgate et al.,
Phys. Rev. Lett. \textbf{85}, 4972 (2000).}


\bibitem{datahid} {B.M. Terhal, D.P. DiVincenzo and D.W. Leung,
Phys. Rev. Lett. \textbf{86}, 5807 (2001); D.P. DiVincenzo et al.,
IEEE Trans. Inf Theory \textbf{48}, 580 (2002); D.P. DiVincenzo et
al., quant-ph/0207147.}
\bibitem{datahid2} {T. Eggeling and R.F. Werner, Phys. Rev. Lett. \textbf{89}, 097905
(2002).}

\bibitem{Zan01} {P. Zanardi,  Phys. Rev. Lett. \textbf{87}, 077901
(2001).}

\bibitem{Barn} {H. Barnum et al., quant-ph/0207149 and
quant-ph/0305023.}

\bibitem{Wiseman} {H.M. Wiseman and J.A. Vaccaro,
quant-ph/0210002; S.D. Bartlett and H.M. Wiseman,
quant-ph/0303140.}

\bibitem{Wer89} {R.F. Werner,  Phys. Rev. A \textbf{40}, 4277 (1989).}


\bibitem{bc} {H.K. Lo and H.F. Chau,  Phys. Rev. Lett. \textbf{78}, 3410
(1997); D. Mayers, Phys. Rev. Lett. \textbf{78}, 3414 (1997).}

\bibitem{May02} {This can be proven as follows: given two pure states for which Bob's reduced
density operators are equal, then the two pure states are equal up
to a local unitary transformation at Alice's side, even in the
presence of superselection rules. Therefore absolute security with
relation to cheating for Bob implies absolute freedom to cheating
for Alice. See also D. Mayers, quant-ph/0212159.}


\bibitem{IP} {N. Schuch, F. Verstraete and J.I. Cirac, {\em In preparation}.}

\bibitem{note2} A similar effect applies in the {\em  normal} quantum data
hiding protocols, where it was shown that an amount of
entanglement proportional to the dimension of the Hilbert space of
the hiding states is necessary to reveal the secret. Note however
that the scaling behaviour is much more favorable in the present
case, where in principle an infinite amount of entanglement is
needed even when the hiding state is a qubit.


\bibitem{Jetal} {D. Jaksch et al., Phys. Rev. Lett. \textbf{82}, 1975 (1999).}


\bibitem{Myln} {O. Carnal and J. Mlynek, Phys. Rev. Lett. \textbf{66}, 2689 (1991);
D.W. Keith et al., Phys. Rev. Lett. \textbf{66}, 2693 (1991).}

\bibitem{CGZ} {J.I. Cirac et al., Phys. Rev. A \textbf{54}, R3714 (1996); Y. Castin and J. Dalibard
Phys. Rev. A \textbf{55}, 4330 (1997).}


\end{thebibliography}
\end{document}